\documentclass{article}
\usepackage[english]{babel}

\usepackage[letterpaper,top=2cm,bottom=2cm,left=3cm,right=3cm,marginparwidth=1.75cm]{geometry}

\usepackage{amsmath}
\usepackage{graphicx}
\usepackage[colorlinks=true, allcolors=blue]{hyperref}

\usepackage{natbib}
\def\beq{\begin{equation}}
\def\eeq{\end{equation}}
\def\beqa{\begin{eqnarray}}
\def\eeqa{\end{eqnarray}}

\title{Physics-informed neural wavefields with Gabor basis functions}
\author{Tariq Alkhalifah and Xinquan Huang}
\date{King Abdullah University of Science and Technology}
\begin{document}
\maketitle
%\begin{frontmatter}
%\title{Physics-informed neural wavefields with Gabor basis functions}
%\author{Tariq Alkhalifah and Xinquan Huang}

\begin{abstract}
Recently, Physics-Informed Neural Networks (PINNs) have gained significant attention for their versatile interpolation capabilities in solving partial differential equations (PDEs). Despite their potential, the training can be computationally demanding, especially for intricate functions like wavefields. This is primarily due to the neural-based (learned) basis functions, biased toward low frequencies, as they are dominated by polynomial calculations, which are not inherently wavefield-friendly. In response, we propose an approach to enhance the efficiency and accuracy of neural network wavefield solutions by modeling them as linear combinations of Gabor basis functions that satisfy the wave equation. Specifically, for the Helmholtz equation, we augment the fully connected neural network model with an adaptable Gabor layer constituting the final hidden layer, employing a weighted summation of these Gabor neurons to compute the predictions (output). These weights/coefficients of the Gabor functions are learned from the previous hidden layers that include nonlinear activation functions. To ensure the Gabor layer's utilization across the model space, we incorporate a smaller auxiliary network to forecast the center of each Gabor function based on input coordinates. Realistic assessments showcase the efficacy of this novel implementation compared to the vanilla PINN, particularly in scenarios involving high-frequencies and realistic models that are often challenging for PINNs.

\vspace{0.1in}

\end{abstract}
%\end{frontmatter}

% Head 1
\section{Introduction}

Wavefield simulation is a fundamental ingredient to many scientific disciplines and applications, ranging from seismology and nondustrictive testing to medical imaging and ultrasonics. Accurate and efficient wavefield simulation plays a pivotal role in understanding the wave propagation phenomena and extracting valuable information about the underlying medium's properties. However, despite the significant advancements in numerical methods that took place over the years, there remain several challenges in the current approaches \cite[]{ELSAYED2004763}. Numerical methods, such as finite difference, finite element, and spectral methods, often encounter difficulties in dealing with complex boundary conditions, large-scale problems, and irregular data observations. Additionally, they may suffer from numerical errors or encounter prohibitive computational costs for real-time simulations or large-scale three-dimensional scenarios. Addressing these limitations is crucial to enhancing the reliability and practicality of wavefield simulation, enabling a deeper understanding of natural and engineered wave phenomena and contributing to various scientific and industrial applications.

Recently, machine learning (ML) and, specifically, deep neural networks (DNNs) have made tremendous inroads in helping us solve physical problems \cite[]{BERG201828,Sorteberg2018ApproximatingTS}.
Considering that DNNs can act as universal function approximators \cite[]{pinkus_1999}, and considering that automatic differentiation (AD), which is available in almost all ML packages \cite[]{baydin2015automatic}, provides pseudo-functional derivatives with respect to the input,
\cite{RAISSI2019686} demonstrated the network's ability to learn functional solutions to nonlinear partial differential equations,
and referred to this process as physics-informed neural networks (PINNs). The functional representation of the PDE solution provides considerable 
flexibility in handling irregular domains and providing compact solutions \cite[]{LLANAS2006283,9933886}.
In spite of these features, PINNs are still trying to carve out their place as a credible alternative to numerical methods, as one main obstacle still persists, especially for wavefields, is their high training cost.

For wavefield solutions, especially those satisfying the Helmholtz equation with inhomogeneous parameters, which are often needed for seismic and ultrasound applications \cite[]{Cuomo}, PINNs have been vigorously tested \cite[]{10.1093/gji/ggab010,10.1093/gji/ggac399,ALKHALIFAH202111,doi:10.1190/geo2022-0479.1}, and many improvements to the vanilla PINN implementation have been suggested. These improvements include modifications to the activation function \cite[]{sitzmann2020implicit,alsafwan2021time}, improvements in the input representation \cite[]{huang2021a,huang2023efficient}, improvements in incorporating boundary conditions \cite[]{2022,9897269,Harpreet}, complementing the loss function with adaptive weights \cite[]{DING2023106425,Shukla2021APN}, improvements in handling high frequencies \cite[]{https://doi.org/10.1029/2021JB023703} and so on. However, with all these improvements, the cost of obtaining high-frequency neural network wavefield solutions is still high, much higher than those obtained using conventional numerical methods \cite[]{doi:10.1190/geo2017-0618.1}. This will limit their use in imaging and inversion applications \cite[]{9585726,https://doi.org/10.1029/2021JB023120}. A big reason for that is the inherent low-frequency bias of neural networks \cite[]{neal2019modern}. In training the neural network to fit a function, it often learns the low-frequency/long-wavelength components of the solution much faster than the high-frequency components. So, the number of epochs needed to attain a solution that satisfies the Helmholtz for high frequencies can be high.

In this paper, we aim to improve the efficiency and accuracy of PINNs for such wavefield solutions by formulating the solution as linear combinations of Gabor functions that satisfy the wave equation. Gabor functions \cite[]{Gabor1946TheoryOC} have proven to be optimal basis functions for wavefields \cite[]{pinto_kolundžija_vetterli_2014}, especially for cases where the wavefield wavelength  (i.e. the wave velocity) changes with space. The Gaussian localized nature of Gabor functions allows for a variable wavelength representation of the solution in space, which is an ingredient needed by PDEs with variable parameters. So, here, we train the network to predict wavefield solutions as a summation of learnable Gabor functions that satisfy the Helmholtz equation. In the examples, we will demonstrate that this implementation alone provides considerable accuracy and efficiency improvements over the conventional implementation. The contribution of this paper can be summarized as follows:
 \begin{itemize}
 \item The utilization of Gabor basis functions that satisfy the underlying wave equation in physics-informed neural networks.
 \item The Gabor function includes parameters that are learned as part of the PINN training, like direction and width.
 \item The utilization of another smaller auxiliary network to predict the location of the center of the Gabor function to allow the learned Gabor functions to contribute to many locations in the solution space.
 \item The application of the approach to realistic models to obtain relatively high-frequency wavefields compared to what has been published previously.
 \end{itemize}
 
The structure of this paper is as follows. We commence by reviewing related work, followed by presenting the form of the Helmholtz equation we aim to solve, which defines the loss function for physics-informed neural networks (PINNs). We next share the Gabor function form used here and the parameters that control their shape, which leads us to the description of their inclusion into PINNs, which is implemented in a multiplicative fashion. We dedicate a section to describe an auxiliary network tasked with predicting the center of the Gabor functions, which constitutes a novel form of PINN that can be useful for applications beyond wavefield modeling. We follow that up with first a simple numerical example to compare its performance with the vanilla PINN, and then we show results for a more realistic model and higher frequencies. The paper concludes by outlining both the strengths and limitations of this approach in the discussion section, culminating in a conclusive summary.
 
\subsection{Related work}
 
Utilizing the Gabor function in neural networks has often enhanced its performance, and this enhancement is most prevalent in convolutional neural networks (CNNs). These functions, inspired by human visual perception, are adept at capturing both local and global patterns in data.  Moreover, the neural encoding properties of Gabor filters align with certain aspects of human visual and auditory perception, rendering them advantageous in modeling systems \cite[]{Olshausen1996EmergenceOS}. The integration of Gabor functions into neural network architectures, thus, offers a synergistic blend of computational efficiency and enhanced modeling capabilities, establishing them as valuable tools for advancing various scientific and technological domains. In image analysis, Gabor filters are employed for feature extraction due to their ability to efficiently capture texture and orientations in images \cite[]{976781,1227801}. By convolving input data with Gabor filters, neural networks can detect complex structures and textures, facilitating tasks such as image classification, object recognition, and segmentation \cite[]{HU20201116,wang2023learnable}. Additionally, Gabor functions' inherent translation and rotation invariant features make them valuable in scenarios where data may exhibit geometric variations \cite[]{YAO202222}. Though the predominant use of Gabor filters is in CNNs, multiplicative activation functions allow for the injection of any function into the MLP architecture \cite[]{fathony2021multiplicative}. This is important since MLP forms the backbone of the physics-informed neural network implementation \cite[]{RAISSI2019686}. In fact, \cite{huang2023gaborpinn} used Gabor functions as a multiplicative activation function for a PINN implementation and noticed improved convergence as it allows for larger imprint of the input coordinates in the network training, like positional encoding. However, they also realized that properly initialing the wavelength of the plane wave within the Gabor function was crucial to its stability. We take this work one step further and include Gabor functions, not just as simple multiplicative activation functions, but solutions to the underlying PDE. This implementation required innovative modifications to the PINN architecture that will be shared here.

\section{Methodology}

We begin by providing an overview of the original Physics-Informed Neural Networks (PINNs) and then delve into the specific Partial Differential Equation (PDE) we are tackling, a variant of the Helmholtz equation which solves for the scattered wavefield. We share our rationale for selecting this particular formulation of the Helmholtz equation and outline the inherent challenges that arise when applying conventional PINNs to address it. Subsequently, we introduce the Gabor function as the foundational basis for solving the Helmholtz equation within our network architecture. We offer an explanation of how these Gabor functions are incorporated into the network. Lastly, in the following section, we present a network modification aimed at optimizing the Gabor-based PINN. This modification enables us to achieve improved accuracy while utilizing fewer Gabor neurons, thus reducing the computational cost.

\subsection{Physics informed neural networks}

Physical phenomena, like wavefields, are often mathematically described as functions of time and space, satisfying partial differential equations (PDEs) that constrain their behavior. Numerically, these functions are represented in discrete form on a particular mesh within a finite domain. However, recently, Multilayer Perceptrons (MLPs) are showing their effectiveness in representing these physical phenomena as functions, considering their function approximation properties \cite[]{pinkus_1999}, especially those differentiable functions describing physical states. Thus, MLP is renowned not only for their remarkable capabilities in learning complex relationships in data, but also for their capacity to serve as universal function approximators \cite{HORNIK1989359}. MLPs orchestrate a cascade of computations within each layer comprising neurons that engage in a weighted aggregation of inputs followed by the application of nonlinear activation functions. The general form of operations within a hidden layer for the output of a neuron, is given by
\beqa
h_i = \sigma \left(\sum_{j=1}^{N_{i-1}} w_{j,i} h_{j,i-1} + b_i \right), 
\label{eq:MLP}
\eeqa
where $N_{i-1}$ signifies the number of neurons in the previous layer $i-1$, $ h_{j,i-1} $ denotes the input feature from the $j$th neuron of the $i-1$ layer, $ w_{j,i} $ represents the weight associated with the input $h_{j,i-1}$, $ b_i $ signifies the bias term for output neuron, and $ \sigma(\cdot) $ embodies a chosen activation function. For PINNs, the last hidden layer connects linearly to the output layer, no activation functions are used. The culmination of these operations across the layers enables the MLP to embrace relationships that extend beyond the confines of the linear approximation. These MLP functions were often used to learn data and functions in a supervised manner. 

With the power of automatic differentiation \cite[]{baydin2015automatic}, \cite{RAISSI2019686} managed to introduce a generally unsupervised form of training the MLP function by using the PDE as a term in the loss function. So, the essense of PINNs lies in the formulation of a loss function that combines two key components: the data misfit term and the physics regularization term. The data misfit term quantifies the discrepancy between the network's predictions and the observed data, while the physics regularization term ensures adherence to the underlying PDE. To train the network, often collocation points within the solution domain are strategically chosen to ensure that the network's predictions adhere to the governing physics equations. These points are typically located at various positions within the domain of interest, often including the boundary and initial conditions locations, where the PDE should be satisfied. The choice of collocation points aims to capture the system's behavior accurately while enforcing the fundamental physical relationships. 
After training the network, which amounts to solving the boundary value PDE problem, we end up with a neural network model that hopefully can predict the solution for any input coordinate value from the solution domain and potentially beyond, acting like an actual function. Conventionally, we evaluate the PINN after training on a regular grid to compare the solution with those evaluated numerically on a regular grid.

\subsection{The PDE}

Solving the Helmholtz equation is crucial to many applications, including seismic and ultrasound imaging and waveform inversion \cite[]{gilbert-kawai_wittenberg_2014}. The classic Helmholtz equation for an isotropic medium with wave velocity $v(\mathbf{x})$, as a function of Euclidean space coordinates, $\mathbf{x} = \{x, y, z\}$, is given by
\beq
\left(\nabla^2+ k^2(\mathbf{x}) \right) u (\mathbf{x})=f(\mathbf{x}), \,\,\, \text{where}  \,\,\,  k(\mathbf{x}) = \frac{\omega}{v(\mathbf{x})}.
\label{eqn:eq1}
\eeq
The solution of this equation is a complex-valued wavefield, $u=\{u_r, u_i\}$, which is a function of the angular frequency, $\omega$, and space coordinates. 
The operator $\nabla^2$ represents the Laplacian, which is the dot-product of two gradient operations, $\nabla$. Finally, $f$ represents the source function, which for many applications, like determining the Green's function, is given by a point source, or in other words. a Dirac delta function. This type of source can be challenging to incorporate in many numerical methods for solving the Helmholtz equation, but in particular, in PINNs \cite[]{ALKHALIFAH202111}.

To address this issue with PINNs, \cite{eage:/content/papers/10.3997/2214-4609.202010588} proposed to include the source function in a background homogeneous medium (constant wave velocity, $v_0$) in which the wavefield solution, $u_0$, can be evaluated analytically. Thus, we solve for the
remaining, often called scattered, wavefield, $\delta u=u-u_0$, using PINNs. In this context, with the residual (perturbation) velocity given by $\delta m=\frac{1}{v^2}-\frac{1}{v_0^2}$, the scattered wave equation has the following form:
\begin{eqnarray}
	\left(\nabla^{2} + { \omega^2 \over v^2} \right) \delta u = -\omega^{2 }\delta m \, u_{0}.
	\label{eqn:eq2}
\end{eqnarray}
This equation can also be referred to as the Lippmann–Schwinger form of the wave equation \cite[]{PhysRev.79.469,7d907e4319894b8c8d879b74cd1f04c2}. This equation is exact as we do not apply the Born
approximation.
For the scattered wavefield, the original (point) source is absorbed by the analytical background wavefield, and the resulting source function in equation~\ref{eqn:eq2} now depends on the perturbation model, which may extend the full space domain.

\subsection{Gabor filters as basis functions}

Gabor basis functions are a set of complex-valued functions used in signal processing and image analysis. They are widely employed for tasks such as feature extraction, texture analysis, and image representation. Gabor functions are well-known for their ability to capture both local spatial information and frequency characteristics of signals and images. They are defined in the spatial domain as a product of a Gaussian envelope and a complex sinusoidal plane wave, which allows them to be tuned to specific frequencies and orientations and thus, Gabor functions have been effective basis functions for complex wavefields, including for seismic data \cite[]{Arnold:95}. The Gabor basis functions are commonly used in various applications, such as image denoising, edge detection, and texture classification.

\begin{figure}[h]
  \centerline{\includegraphics[width=8cm]{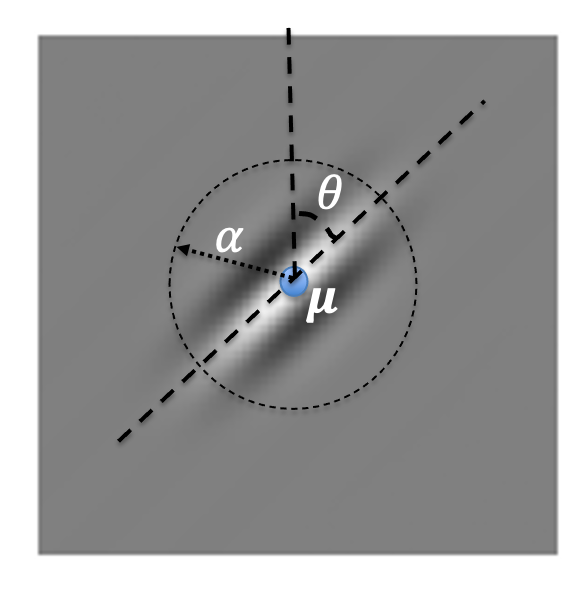}}
  \caption{An example Gabor function of the plane wave with angle, $\theta$, wavelength, $\lambda$, and weighted by a Gaussian function with a mean located at ${\bf \mu}$, and variance, $\alpha$.}
  \label{fig1}
\end{figure}

A Gabor function in 2D, represents a plane wave with a specific wavenumber ($k$), a phase shift ($\phi$), and direction ($\theta$) weighted by a Gaussian function centered at location (${\bf \mu}$) in the domain of interest with a specific variance ($\alpha$),
 to localize the plane wave. Such a function, with wavenumber $k$,  satisfies the wave equation~\ref{eqn:eq1}. For 2D media, specified by $x$ laterally and $z$ in depth, the Gabor function can be given by: 
\begin{equation}
G(k,x,z) = e^{-\frac{ (\tilde{x}-\mu^x)^2+\beta (\tilde{z}-\mu^z)^2}{\alpha}}  e^{\left(i (k \tilde{x}+\phi)\right)},
\label{eq:2}
\end{equation}
where $i$ is the imaginary unit in this equation and
\begin{eqnarray}
\tilde{x} &=& x \cos\theta + z \sin\theta, \nonumber \\
\tilde{z} &=& z \cos\theta - x \sin\theta.
\label{eq:2_xz}
\end{eqnarray}
Here $\mu^x$ and $\mu^z$ are the $x$ and $z$ components of Gabor center, ${\bf \mu}$, respectively. The parameter, $\beta$, allows for an anisotropic Gaussian weight. Figure~\ref{fig1}  shows a simple plot that describes the role of the various parameters of the Gabor function. In 3D, the direction of the plane wave is described by two angles, with the other angle often representing the azimuth.

Since we want to maintain the wave-equation solution property of Gabor functions, we incorporate them (equation~\ref{eq:2}), in a multiplicative fashion, into the final, $L$, hidden layer of the fully connected neural network where $L$ represents the number of hidden layers, and the connection between the $L$ layer and the output is a mere summation (linear). For the real part of the wavefield, we sum the weighted real part of the Gabor functions (the cosine term), and for the imaginary part, we sum the weighted imaginary part of the Gabor functions (the sine term). So, the input to the this $L$ layer from the previous layers is given by equation~\ref{eq:MLP}, and form the weights/coefficients of the Gabor functions.
Specifically, the output of the $L-1$ layer are multiplied by $w_{j,L}$ with an added bias value, and passed through an activation function, to provide $h_{L}$, which serves as a coefficient (amplitude) for the Gabor function, $G$, within each neuron in the $L$ layer. In mathematical form, the output of a neuron, $y_{L}$, in the Gabor layer, 
$i=L$ (equation~\ref{eq:MLP}), is given by
\beqa
h_{L} &=& \sigma \left( \sum_{j=1}^{N_{L-1}} w_{j,L} h_{j,L-1} + b_{L} \right),  \nonumber \\
y_{L} &=&    h_{L} G.
\label{eq:Gabor}
\eeqa
Finally, the outputs of the neural network, $\delta u_r$ and $\delta u_i$, are mere summations of the real and imaginary parts, respectively, of all the neurons, $y_{j,L}$, in the last layer. This implies that the output wavefield is a linear summation of Gabor functions. The wavenumber of these Gabor (neuron) functions is fixed and given by the
\begin{equation}
k=\frac{\omega}{v(x,z)},
\end{equation}
where $\omega$ represents the frequency we are solving for, and $v$ is the velocity corresponding to the input sample coordinates in 2D $\{x,z\}$. Thus, the input coordinates to the NN function also contributes to the Gabor layer through the velocity and through equation~\ref{eq:2}. Including the frequency in the Gabor function is important for a speedy convergence of PINNs \cite[]{huang2021single} for high frequencies.

In the initial stages of testing our new Gabor-layer network, our primary emphasis is on employing the PDE loss. In this context, our primary focus is on the network's training. This involves utilizing the network to evaluate both the wavefield and its second-order partial derivatives with respect to variables $x$ and $z$. Thus, to train the network, with equation~\ref{eqn:eq2}, we use the following loss function:
\begin{eqnarray}
f=\frac{1}{N}\sum_{j=1}^{N}  && \left | \omega ^{2}m^{(j)}\delta u_r^{(j)}+\nabla^{2}\delta u_r^{(j)}+\omega ^{2}\delta m^{(j)} u_{r0}^{(j)} \right |_{2}^{2} + \nonumber \\
										    &&  \left | \omega ^{2}m^{(j)}\delta u_i^{(j)}+\nabla^{2}\delta u_i^{(j)}+\omega ^{2}\delta m^{(j)} u_{i0}^{(j)} \right |_{2}^{2},
\label{eqn:eq3}
\end{eqnarray}
where $N$ is the number of training samples, and $j$ is the training sample index. The two terms in the loss function correspond to the losses for the real ($\delta u_r$) and imaginary ($\delta u_i$) parts of the scattered wavefield,
using the real ($u_{r0}$) and imaginary ($u_{i0}$) 
parts of the background wavefield. For this loss function, we chose the background model to be simple enough (homogeneous) so that the background wavefield can be evaluated analytically on the fly.

\begin{figure}[h]
  \centerline{\includegraphics[width=18cm]{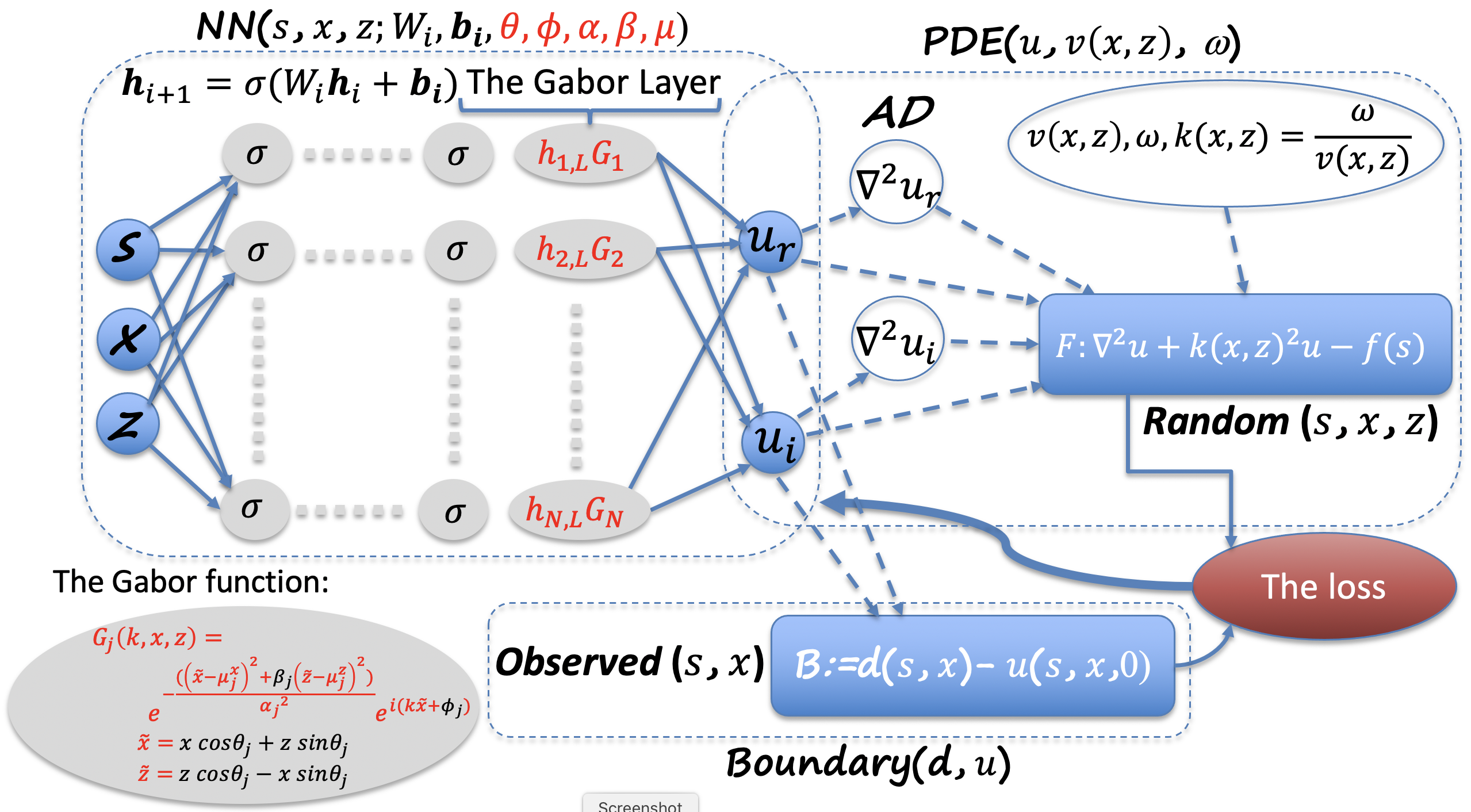}}
  \caption{A schematic diagram of the new PINN architecture with Gabor basis functions. The framework includes three parts, the network NN, and the PDE loss, with its parameters, and the boundary condition. The network takes (for training) random input source locations at the surface, $x_s$, as well as domain coordinate points, $\{x,z\}$, and outputs the real ($u_r$) and imaginary ($u_i$) parts of the wavefield that should satisfy, after training, the Helmholtz equation $F$, and the boundary condition $B$. The neurons of every layer are like the conventional PINN ones with activation functions, denoted by $\sigma$, other than the last layer that includes Gabor functions multiplied by the output of previous layer, with neural network weights $W$ and biases ${\bf b}$. Also learned are the Gabor parameters other than the frequency. The Gabor function $G$ is shared in lower left. AD stands for automatic differentiation. All parameters in the neural network function, $NN$ prior to the semi-column are inputs, and all parameters after are learnable.}
  \label{fig2}
\end{figure}

In summary, the PINN network is given by fully connected hidden layers with activation functions in every layer other than the last. So, to adapt PINNs to wavefields, we multiply the neuron evaluations at the last layer with learnable Gabor functions in which the frequency is that of the Helmholtz equation, and thus, these Gabor functions satisfy the wave equation for that frequency. These Gabor functions, with amplitudes given by the neurons output from the previous layers, are summed to provide the output wavefield values. Thus, the input to the network, like a function, is a location in space, given in 2D by $x$ and $z$ coordinate values, and in 3D by $x$, $y$, and $z$ coordinate values. 
The output of the network consists of the real and imaginary parts of the complex-valued scattered wavefield at the input location. Figure~\ref{fig2} shows, in detail, this PINN network for the case in which we also include the source location on the surface as input, like a Green's function.

\section{The Gabor function Apex/Center location}

To construct a Gabor basis, multiple Gabor functions are typically utilized with different orientations and scales to cover various spatial frequencies and orientations of a predicted wavefield. The Gabor basis functions provide a multi-scale and multi-orientation representation of the wavefield. These scales and orientations are guided by the Gabor function parameters, which are learned as part of training PINNs.

\begin{figure}[h]
  \centerline{\includegraphics[width=8cm]{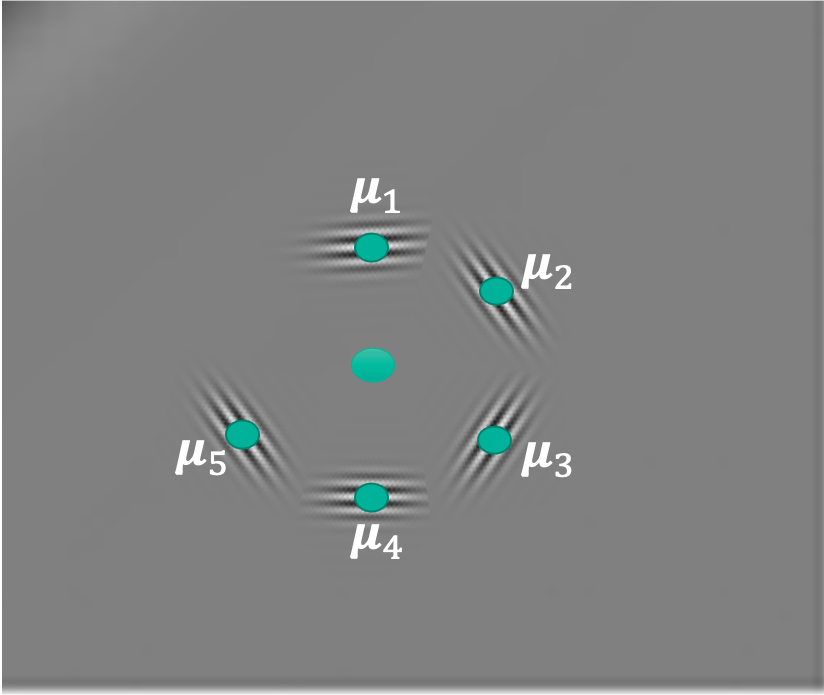}}
  \caption{Gabor functions plotted schematically to show that many of them are required to properly reconstruct the wavefield corresponding to a source in the middle. ${\bf \mu_i}$ are the learned Gabor centers for these five Gabor functions $\{i=1-5\}$. The large dot in the middle represents a source.}
  \label{fig3}
\end{figure}

The multiplicative injection of the Gabor functions into PINNs limits the neurons contributions to an area centered at ${\bf \mu}$, which is a learnable parameter. The size of this area depends on another learnable parameter, the variance, $\alpha$. In other words, after training, the Gabor function becomes stationary in space. If the PDE parameters, like the velocity, changes a lot, the variance of the Gabor function, $\alpha$, is expected to be small to confine the contribution of the plane wave to the area in which its wavenumber reflects the velocity, and we refer to this area as the support region of the Gabor function (The circle in Figure~\ref{fig1}). As a result, in these cases, input coordinate location, $\bf x$, beyond the support region of the Gabor function will render this learned Gabor neuron to have negligible contributions to the solution. This localized by a learned ${\bf \mu}$ contribution is generally fine if the Gabor function was used in every hidden layer and every neuron as conventional multiplicative activation functions. However, for our implementation here, where the Gabor functions satisfy the wave equation and reside in one layer as basis functions, such localized contributions will force us to use too many neurons in the Gabor layer.
Figure~\ref{fig3} highlights this problem in a schematic plot showing the contribution of five Gabor functions that might be needed to build the wavefield emanating for a source in the middle. After training, ${\bf \mu}$ for each Gabor neuron is fixed, and thus, these Gabor functions are stationary. 

 Ideally, we want every Gabor neuron to contribute to the solution space everywhere, which will reduce the number of neurons needed in the Gabor layer.  This will allow a neuron with a specifically learned angle to contribute to all the wavefield that might have waves traveling at that angle, with wavelength computed implicitly from the frequency and the velocity at the input locations. For example, if we had waves emanating from a source in the middle, having a quasi-circular shape, we would expect the Gabor neurons, let us say 50 of them in the layer (this number depends on the wavelength and radius), to cover the plane wave angles, $\theta$, needed to reproduce the circular shape at a certain radius (like Figure~\ref{fig3}). However, for another radius, we will need another batch of Gabor neurons covering practically the same angle distribution to reconstruct the wavefield there. If there were scattering, and thus additional waves, we will need even more neurons to represent the scattering and this can lead to very wide Gabor layer with maybe 1000s of neurons.

\begin{figure}[h]
  \centerline{\includegraphics[width=14cm]{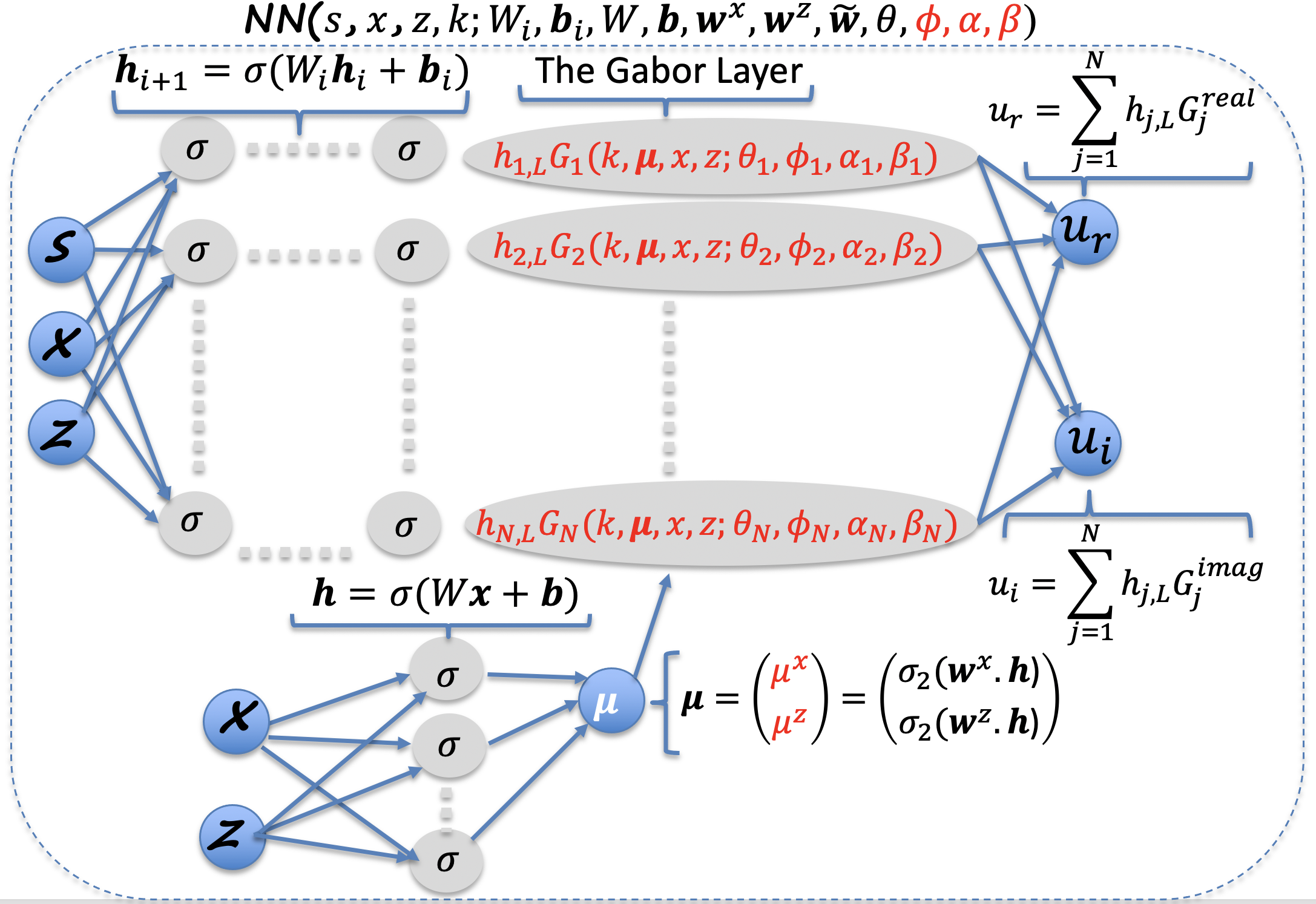}}
  \caption{A diagram of the modified PINN network to that in Figure~\ref{fig2}, which includes auxiliary connections with one hidden layer to predict a single Gabor center, ${\bf \boldsymbol{\mu}}=\{\mu^x,\mu^z\}$, to be used by the neurons in the Gabor layer. The activation function $\sigma_2$ stands for the Sigmoid to insure that the prediction of Gabor center location is within the domain of interest. The vectors ${\bf w^x}$ and ${\bf w^z}$ are the columns of the weight matrix connecting the hidden layer to the two outputs, and the input to the auxiliary connections are the same ${\bf x}=\{x,z\}$ as the main network. All parameters in the neural network function, {\it NN}, and the Gabor function, $G$, prior to the semi-column are inputs, and all parameters after are learnable.}
  \label{fig4}
\end{figure}

To combat this problem, we devise an approach that allows learned Gabor neurons to contribute to the full space domain, regardless of their Gaussian support region. We specifically employ another, much smaller, auxiliary network to learn ${\bf \boldsymbol{\mu}}$ as a function of input ${\bf x}=\{x,z\}$. As a result, the locations of these Gabor functions are dynamically dependent on the input coordinates, and specifically the training samples, and thus, a Gabor function possessing a plane wave angle of 30 degrees (learned), can contribute to many locations in the domain, as many as we have sample points, and we typically have many of those (1000s). Accordingly, the modified architecture of the Network is revealed in Figure~\ref{fig4}. The auxiliary network accepts the space coordinates as input and predicts a single Gabor center ${\bf \boldsymbol{\mu}}$ ($=\{\mu_x,\mu_z\}$, in 2D). In our examples, we use one hidden layer for this auxiliary network, and compute the output with a Sigmoid activation function. This activation function will insure that the predicted center falls within our normalized (between zero and one) space coordinates. The single predicted ${\bf \mu}$ is utilized in all the Gabor neurons within the Gabor layer. Thus, for every training sample, we will have a different Gabor center for the neurons to use, which for the sake of building the wavefield, will most likely be within the vicinity of the input sample, allowing the input sample, considering the Gabor Gaussian window, to contribute. We also experimented with predicting, in the auxiliary network, as many Gabor centers as we have Gabor neurons, but it was not as robust as the single center.

The specific details of the fully connected deep network will be shared in the examples. However, for all the examples, the activation function between layers, other than the last, is the sine function. We chose to optimize the loss function using an Adam optimizer, all full-batch, gradient-based optimization algorithm \cite[]{liu1989limited}. 
We train such PINNs using random points from the solution domain, and after training, we usually evaluate the NN at a regular grid for a single source for ease of visualization. Nevertheless, the scattered wavefield, after training, is stored in an NN and can be evaluated anywhere.

\section{Results}

We test our approach first on a simple classic layer-model example for a wavefield frequency of 4Hz. We then share results for the same model, but hike the frequencies to 16 Hz. Finally, we test the approach on a realistic model in which we often avoid applying PINN to. We conduct this test for a frequency of 16 Hz, as well. In these tests, we compare the results of our Gabor-based PINN to those obtained using the Vanilla PINN by using the numerical solution as reference. For higher frequencies, we incorporate positional encoding \cite[]{huang2021a}.

\subsection{A simple model comparison}

\begin{figure}[h]
  \centerline{\includegraphics[width=14cm]{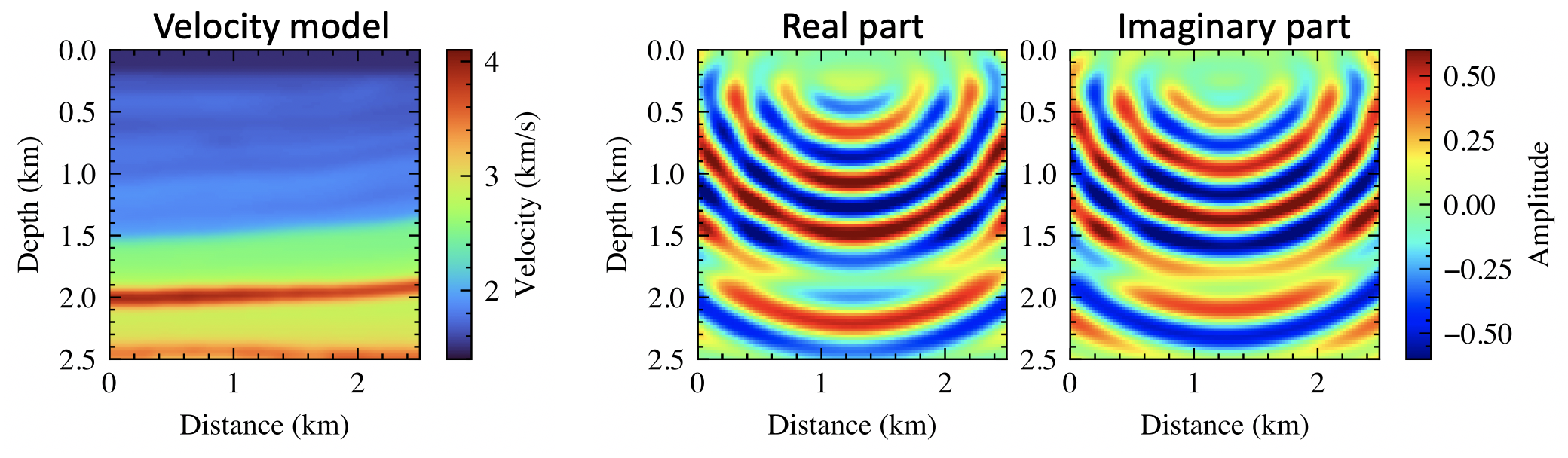}}
  \caption{The velocity model and the real and imaginary parts of the scattered wavefield evaluated numerically by solving the scattered Helmholtz equation~\ref{eqn:eq2} for a frequency of 4 Hz. The background velocity is constant and equal to 1.5 km/s, and the source is up top located in the middle at 1.25 km.}
  \label{fig5}
\end{figure}

\begin{figure}[h]
  \centerline{\includegraphics[width=14cm]{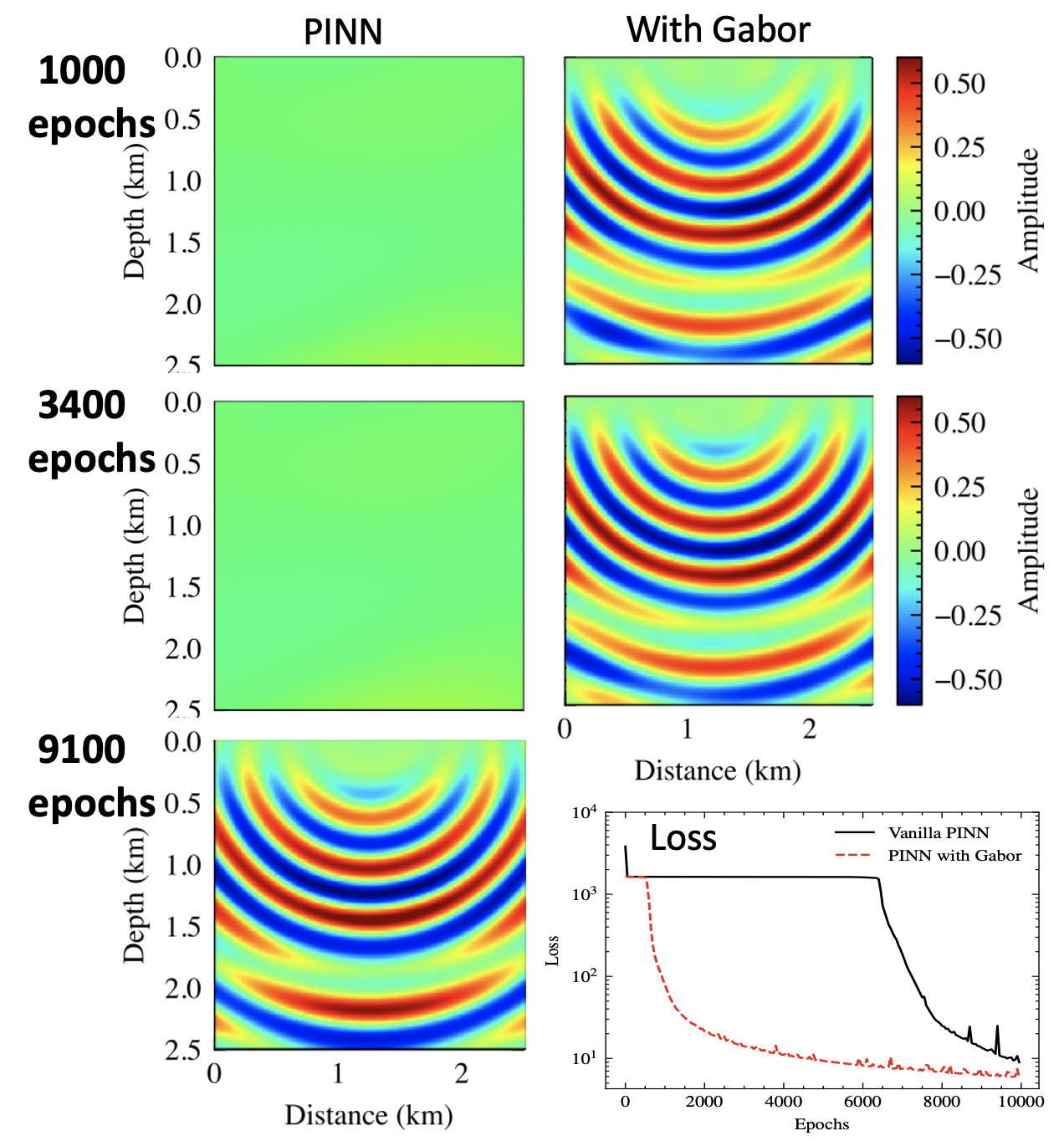}}
  \caption{The predicted real part of the scattered wavefield for the vanilla PINN and the Gabor-based PINN after
  1000 and 3400 (top two rows) epochs. The bottom row displays the predicted real part of the scattered wavefield for the vanilla PINN after 9100 epochs and the loss curves for the vanilla PINN (solid curve) and the Gabor-based PINN (dashed curve).}
  \label{fig6}
\end{figure}

We test the approach initially in solving the scattered Helmholtz equation for the velocity model in Figure~\ref{fig5}, with a background 1.5 km/s velocity, for a 4 Hz wavefield. The real and imaginary parts of the numerical solution for a point source on top at location 1.25 km (the middle) is shown in Figure~\ref{fig5}, as well. Considering that the background velocity represents the velocity of the model up top, we note that the scattered energy is weak there to the point that we avoid the point source singularity. This, as \cite{ALKHALIFAH202111} demonstrated, is helpful to the PINN robustness and convergence. Focusing on the real part of the wavefield, we compare the performance of this new Gabor-based PINN to the vanilla implementation of PINN. For 4 Hz, we use a simple network given by 5 hidden layers of which we have 32 neurons in each of the first 4 layers to predict the coefficients of the Gabor basis functions. For the fifth layer, we have only 16 Gabor neurons. However, thanks to the Gabor center auxiliary subnetwork, the 16 Gabor neurons will contribute to the whole solution domain. For the vanilla PINN, we have the same configuration, but all with conventional neurons. We train both networks using 5000 random samples of ${\bf x}$=\{$x$, $z$\}, and, as we mentioned earlier, we use an Adam optimizer, and a learning rate of 0.001. For the Gabor-based network, the 5000 samples and 16 Gabor neurons, implies that up to $16 \times 5000$ unique Gabor functions may contribute to the solution.
Figure~\ref{fig6} shows the predicted real part of the scattered wavefields on a regular grid (to compare with the numerical, Figure~\ref{fig5}) for the vanilla PINN and the Gabor-based PINN at various epochs. Also shown at the bottom right corner are the loss curves for the training of the two PINNs. As reflected in the loss curves, the Gabor-based PINN converges much faster than the vanilla PINN thanks to the Gabor basis functions. In fact, for the vanilla PINN, 9100 epochs were barely enough to converge to an acceptable accuracy, part of this is due to the relatively small network used in this example. Larger networks will yield better fitting but at a much higher cost.

\begin{figure}[h]
  \centerline{\includegraphics[width=14cm]{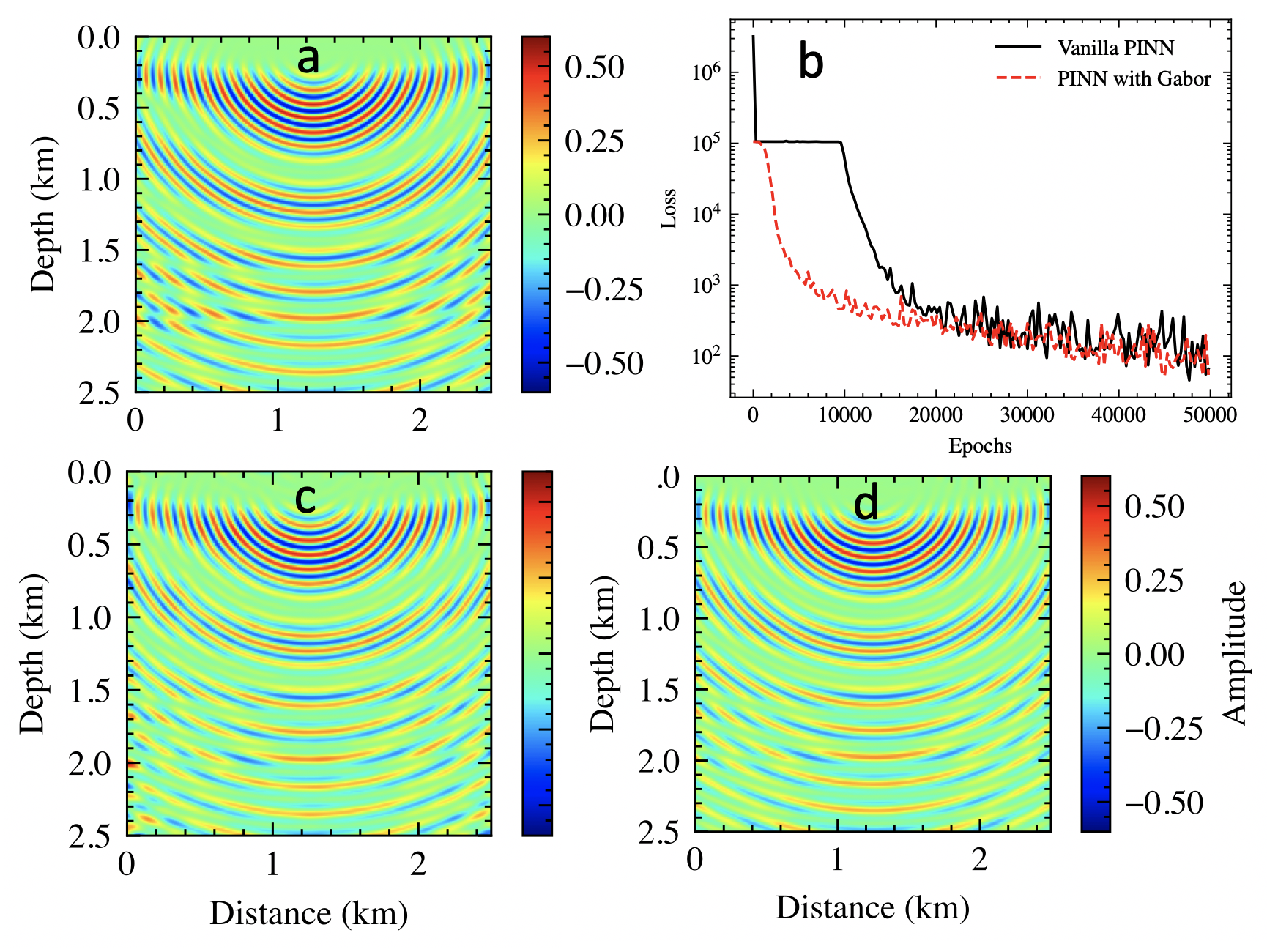}}
  \caption{a) The real part of the scattered wavefield evaluated numerically by solving the scattered Helmholtz equation~\ref{eqn:eq2} for a frequency of 16 Hz. b) The loss curves for the vanilla (solid) and Gabor-based (dashed) PINNs.  The real part of the scattered wavefield from c) the vanilla PINN and d) from the Gabor-based PINN after 50000 epochs.}
  \label{fig5a}
  \end{figure}

\begin{figure}[h]
  \centerline{\includegraphics[width=16cm]{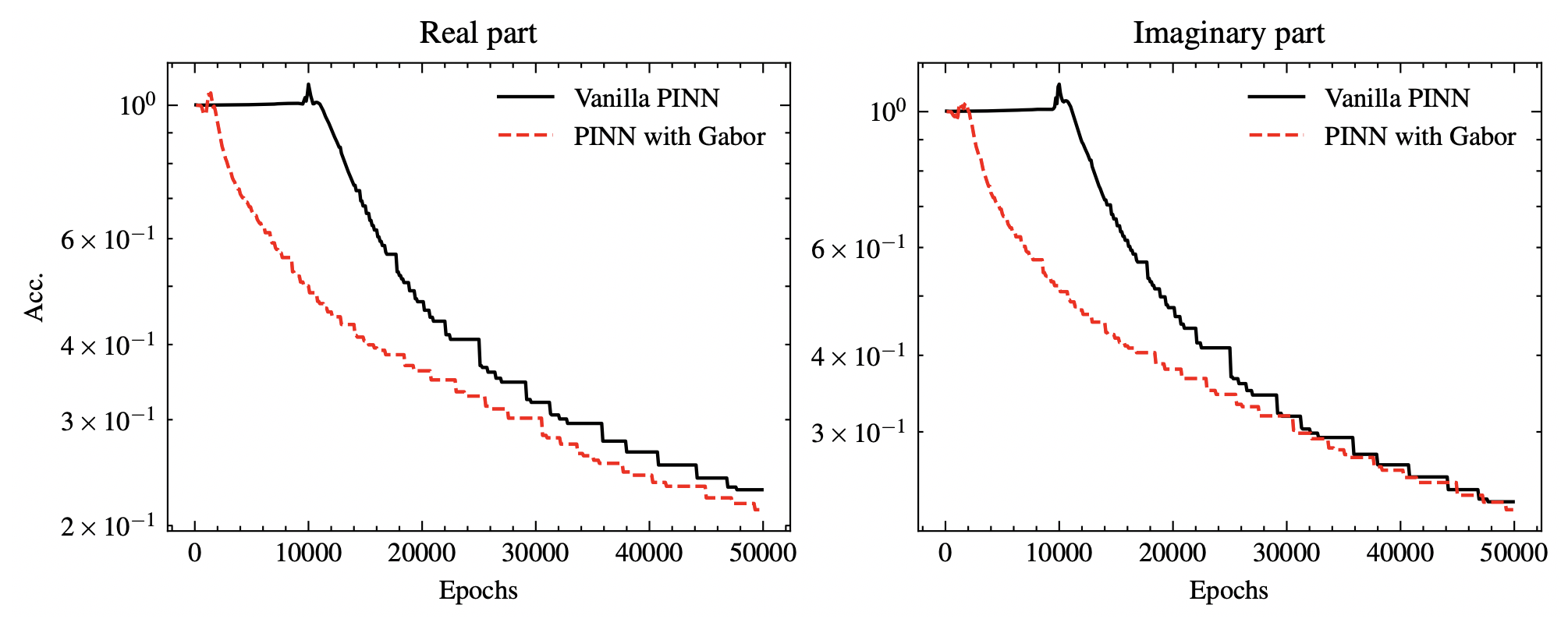}}
  \caption{The mean square error curves for real (left)  and imaginary (right) parts of the scattered wavefield up to 50000 epochs for the vanilla PINN (solid curves) and the Gabor-based PINN (dashed curves).}
  \label{fig6b}
\end{figure}

As mentioned earlier, high frequencies pose a challenge to PINN, as the resulting wavefield solution for the same size domain becomes far more complex to learn. In 2D, the 16 Hz wavefield is effectively about 16 times more complex than the 4 Hz wavefield as we effectively scale each axis by a factor of 4. However, the exact complexity increase depends on the velocity model. We add to both networks (vanilla and Gabor-based) positional encoding that will convert the scalar coordinates input to vectors described by sinusoidals. Without the positional encoding, PINN cannot handle the complexity of the 16 Hz wavefield. The numerical solution of the real part of the scattered wavefield corresponding to the model in Figure~\ref{fig5} is provided Figure~\ref{fig5a}a. We also have to increase the size of the neural network model to $\{128, 128, 64, 64, 64\}$ neurons from shallow to deep. For the Gabor based PINN this implies that we have 64 Gabor basis functions. For the positional encoding, we use $L=4$, which is the dimension of the positional encoding vector for each input variable. The loss curves (of fitting the PDE) for both training are shown in Figure~\ref{fig5a}b. The Gabor-based network converged much faster and had a lower loss after 50000 epochs of training. The resulting real part of the scattered wavefield (Figure~\ref{fig5a}d) for the Gabor-PINN looks closer to the numerical solution than the vanilla PINN solution (Figure~\ref{fig5a}c). This observation is supported by the relative L2 norm error curves (the L2 norm of the differences between the prediction with the numerical solution and then divided by the L2 norm of the numerical solution) for both trainings, shown in Figure~\ref{fig6b}.

\subsection{A more realistic model}

\begin{figure}[h]
  \centerline{\includegraphics[width=14cm]{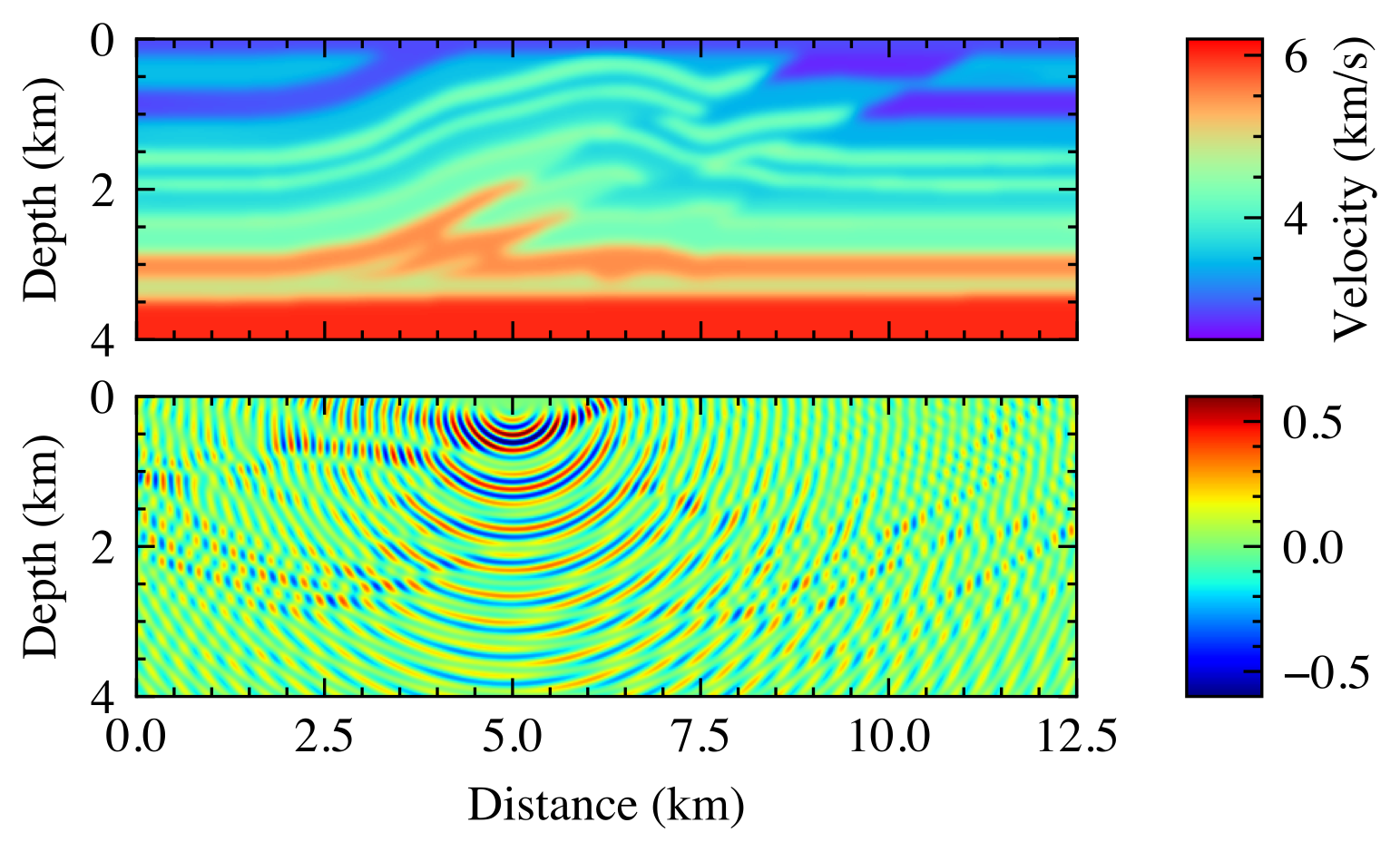}}
  \caption{The velocity model (top) and the real part of the scattered wavefield (bottom) evaluated numerically by solving the scattered Helmholtz equation~\ref{eqn:eq2} for a frequency of 16Hz.The background velocity is constant and equal to 1.5 km/s. The source is located up top in the middle at location 6.25 km.}
  \label{fig7}
\end{figure}

A popular model in the seismic community is what we refer to as the overthrust model, which is supposed to depict the subsurface in the Canadian foothills. The model stretches 12.5 km wide and 4 km deep, and includes many layers, but the most imposing feature in the model are the two overthrust faults in the middle (Figure~\ref{fig7}). The complexity of this model is enhanced by the size of the model, which is 8 times the size of the previous model, and yet we will solve for a 16 Hz wavefield. The numerical solution for the real part of the scattered wavefield is shown in Figure~\ref{fig7}, and we can probably appreciate the complexity of the solution we are asking PINN to learn. The number of wave cycles within the model is large. We also have used a far more complex model that induces more scattering. As a result, we use a larger network of 5 hidden layers with \{256, 256, 128, 128, 128\} neurons per layer from shallow to deep. Note that, in this example, we will have 128 Gabor basis functions contributing to the solution. To help PINN converge for such a problem, we again equip both PINNs (vanilla and Gabor-based) with positional encoding with $L=4$.

\begin{figure}[h]
  \centerline{\includegraphics[width=10cm]{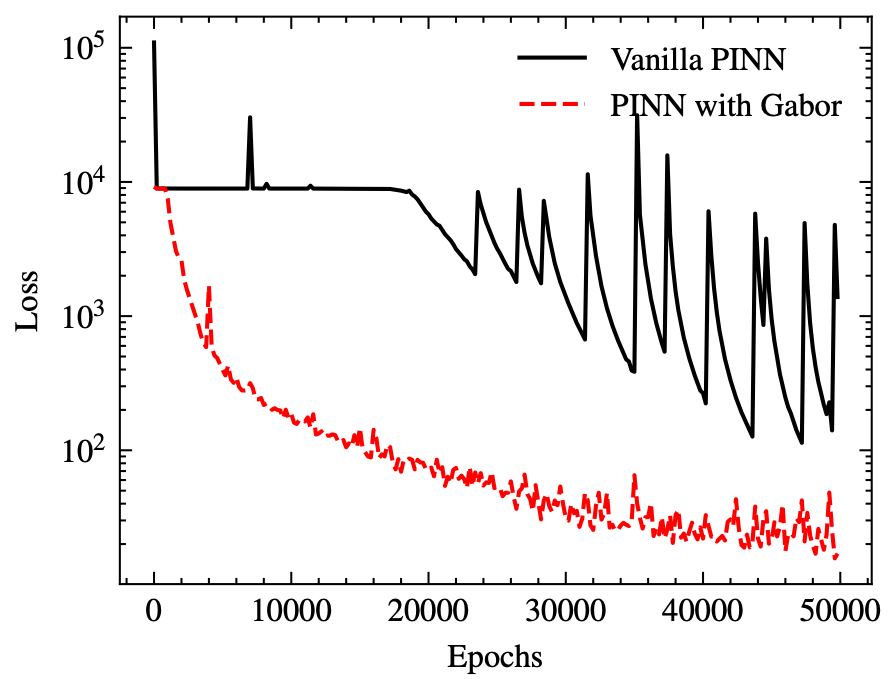}}
  \caption{The loss curves for the vanilla PINN with positional encoding (solid) and the Gabor-based PINN with positional encoding (dashed).}
  \label{fig8}
\end{figure}

The loss curves for both networks (vanilla and Gabor-based) are shown in Figure~\ref{fig8}. The Gabor-based PINN, enabled by Gabor basis functions, exhibits an initial descent in the loss curve, unlike the vanilla PINN, which initially undergoes parameter space exploration to acquire the ability to model sinusoidal functions before converging. The positional encoding was helpful to both PINNs in their convergence, considering the high frequency consideredhere and the relatively large size of the model.

\begin{figure}[h]
  \centerline{\includegraphics[width=14cm]{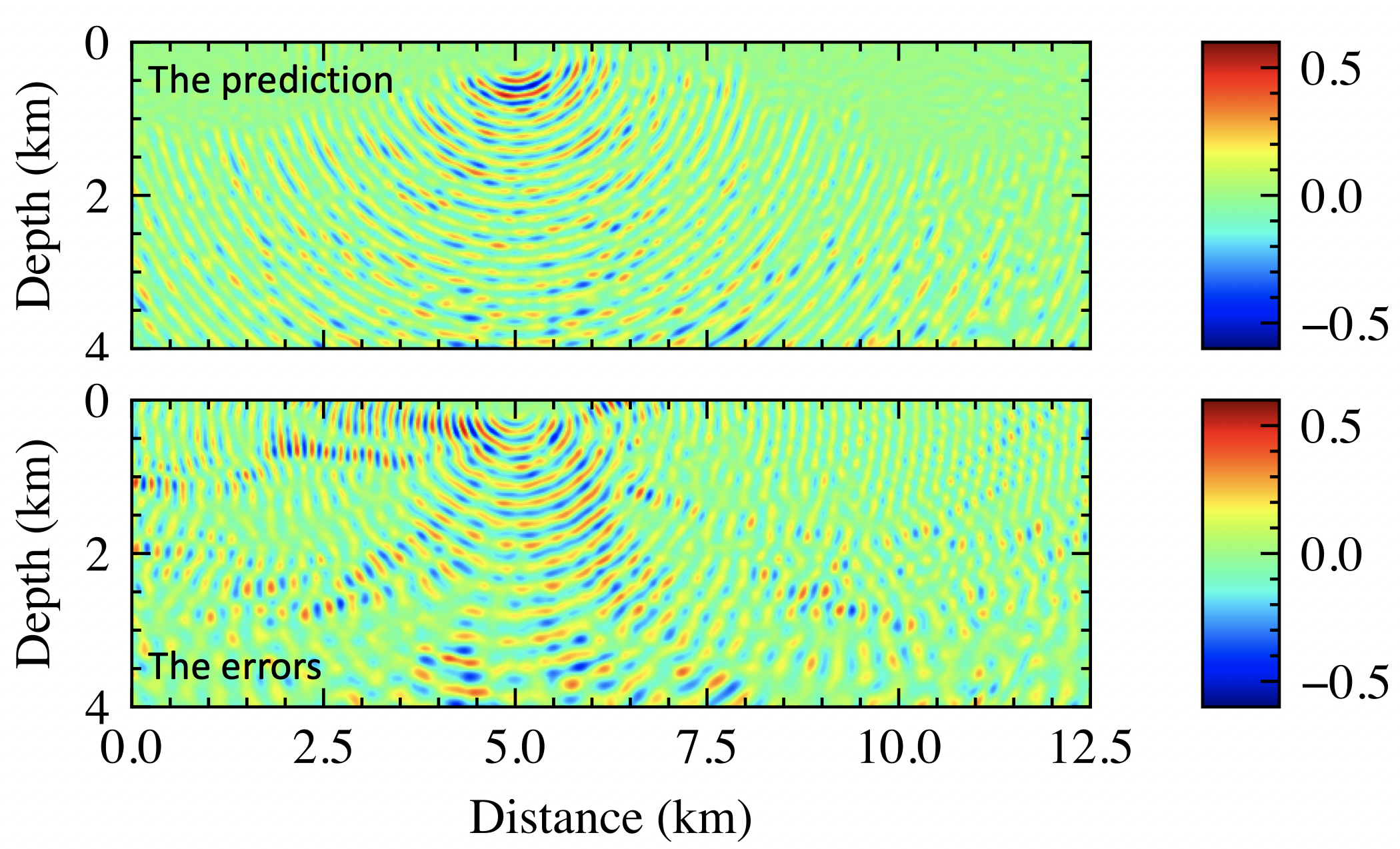}}
  \caption{Top: The predicted real part of the scattered wavefield using the vanilla PINN with positional encoding for the model in Figure~\ref{fig7}. Bottom: the difference with the numerical solution shown in Figure~\ref{fig7}.}
  \label{fig9}
\end{figure}

The solution provided by the vanilla PINN for the real part of the scattered wavefield after 50000 epochs is shown in Figure~\ref{fig9} (top). We also show the errors (difference with the reference numerical solution) at the bottom. With the help of the positional encoding, we managed to somewhat converge, but the errors are relatively large. Actually, the errors have comparable energy to that of the solution. 

\begin{figure}[h]
  \centerline{\includegraphics[width=14cm]{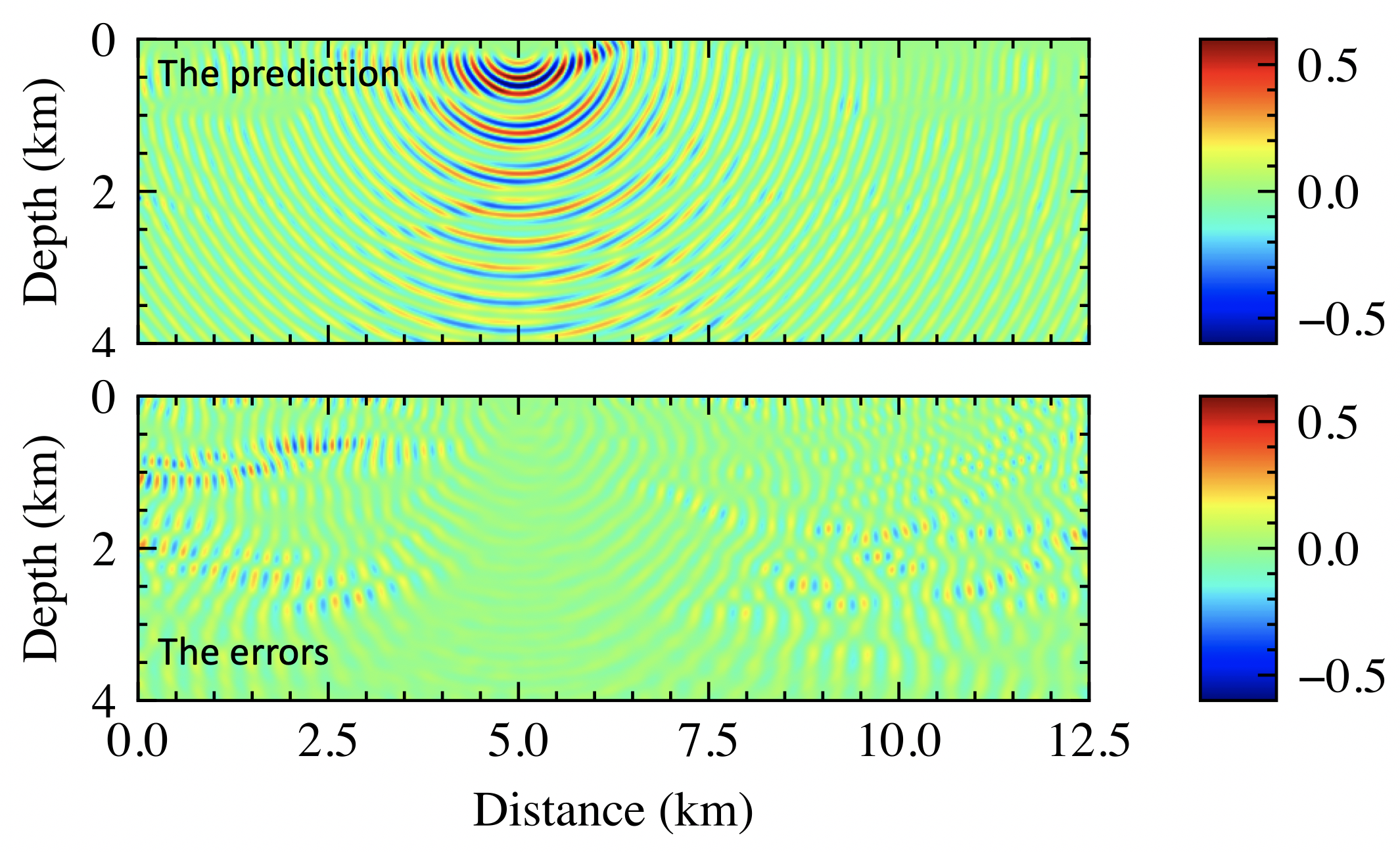}}
  \caption{Top: The predicted real part of the scattered wavefield using the Gabor-based PINN with positional encoding for the model in Figure~\ref{fig7}. Bottom: the difference with the numerical solution shown in Figure~\ref{fig7}.}
  \label{fig10}
\end{figure}

\begin{figure}[h]
  \centerline{\includegraphics[width=14cm]{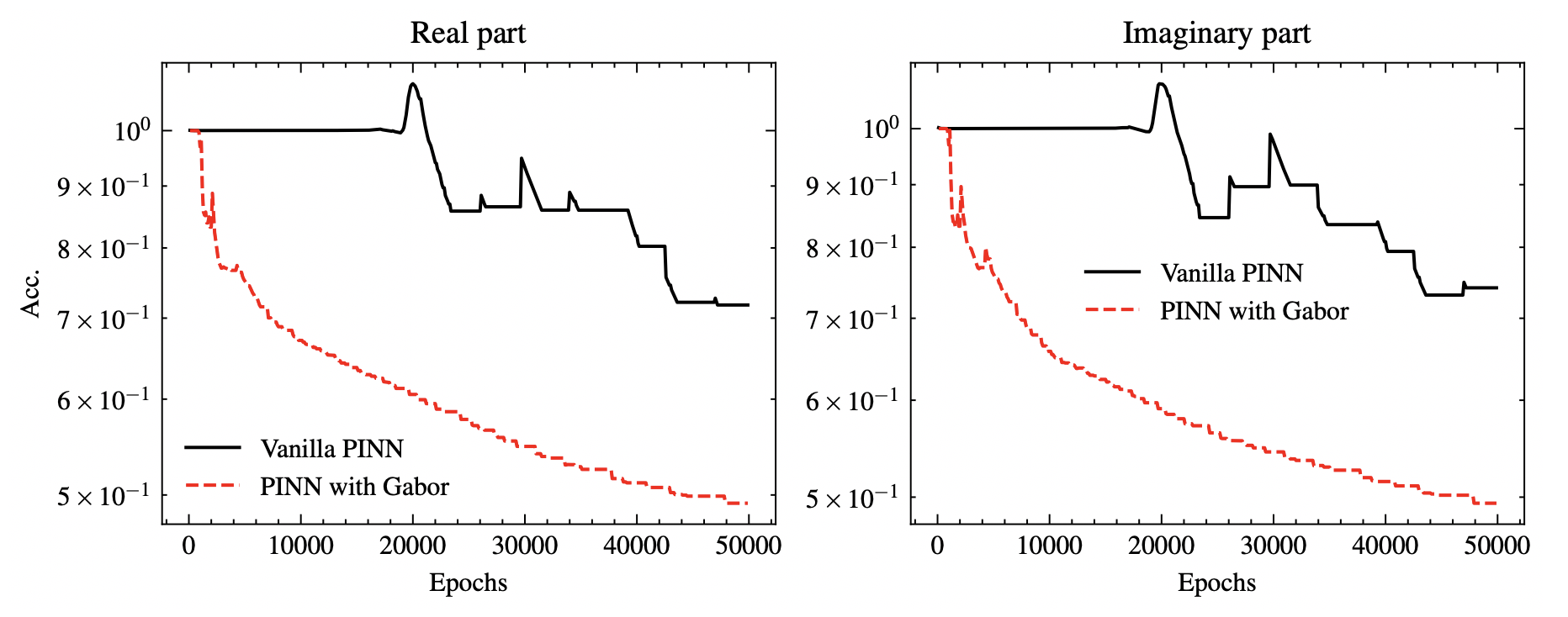}}
  \caption{The accuracy of the prediction for the real part (left) and imaginary part (right) for the vanilla and Gabor-based PINNs.}
  \label{fig11}
\end{figure}

The solution from the Gabor-based PINN for the real part of the scattered wavefield after 50000 epochs is shown in Figure~\ref{fig10} (top). The corresponding errors are relatively low. In fact, by observing the relative L2 norm error between the numerical solution and those of PINN and Gabor-based PINN evolve over epochs in Figure~\ref{fig11}, we notice that the Gabor-based PINN converges much faster to what we deem as the reference solution (The numerical solution).

\section{Discussion}

Generally, Gabor-based neural networks incorporate Gabor wavelet filters as a preprocessing step, aiming to enhance feature extraction from images. However, they might result in increased computational complexity due to the intricate nature of Gabor wavelets, potentially resulting in longer training times and higher resource requirements. These networks might struggle with generalizing effectively across diverse data, as they are tailored to specific localized patterns captured by Gabor filters. On the flip side, Gabor-based PINNs can navigate away from these weaknesses and, more importantly, provide interpretability of learned parameters as they describe the Gabor basis functions, like amplitude, angle, width, and so on.

Multi-layer perceptron (MLP) has an inherent bias to match low frequencies far earlier in the training cycle than high frequencies. This feature is driven by the polynomial nature of the MLP neuron operations, and polynomials, even with the added nonlinearity of activation functions, have usually had a hard time approximating periodic functions like waves.
On the other hand, Gabor functions are by nature sinusoidal and, with the imposed Gaussian weights, are localized (nonstationary). Thus, incorporating Gabor functions into the linear part of the MLP architecture of PINNs allows us to formulate wavefield solutions using Gabor basis functions that satisfy the wave equation. However, the limited support of the Gabor functions renders their contribution to a limited region around the learned center (apex) of the Gabor function. However, if we make the Gabor center a function of the input coordinates through a simple learnable auxiliary network, we
allow the Gabor neurons to contribute everywhere we have samples, with the amount of contribution controlled by the Gabor coefficients corresponding to the predictions of the previous layers.

A sigmoid function is used for the output of the auxiliary network to insure that the predicted Gabor center stays within the desired domain. Of course, we can map the 0 and 1 range of the Sigmoid to any size solution domain. If the domain is irregular, we can map the Sigmoid's range to the far edges of this irregular domain to insure we cover the full domain. In this case, some of the predicted Gabor centers may fall outside the solution domain, but we assume that the auxiliary network would be motivated to have them contribute to the solution as part of the optimization. 

There are numerous avenues for enhancing the efficacy of Gabor functions in contributing to the solution, thereby reducing the reliance on a large number of Gabor functions. One approach is to employ another small auxiliary network, typically also one layer deep, to predict the Gabor plane wave angle for each input sample. Consequently, each Gabor neuron can have a specific angle (slope) based on the sample's location. This will allow us to reduce the Gabor layer width to as small as 4 or 5 neurons, and thus, the alternative network for angles can learn 4 or 5 angles per input sample. Such slopes can often cover all potential wave angles at each input sample location, thus handling up to 4 or 5 crossing waves, respectively. We will still need wide layers prior to this thin Gabor layer to accommodate all the possible coefficients (amplitudes) for the Gabor functions for each input sample. 

%In summary, we utilized the complex Gabor filter as basis functions to help us predict the real and imaginary parts of the wavefield solution. This was accomplished by connecting the output of the cosine term of the Gabor function to the real part and the sine term to the imaginary part of the scattered wavefield solution, as they both share the same weights (the formal definition of a basis function). In previous work \cite{huang2023gaborpinn}, it has been shown that Gabor functions can also serve as multiplicative activation functions utilizing only the real part of the Gabor filter. The convergence of PINN improved considerably with such activation functions. However, we had to be careful in initializing the Gabor frequencies, and this depended on the frequency of the wavefield, though this Gabor implementation does not utilize the Gabor filters as basis functions.

\section{Conclusions}

We equipped physics-informed neural networks (PINNs) with learnable Gabor functions that satisfy the Helmholtz equation. The Gabor functions are placed in the last hidden layer, connected linearly to the output real and imaginary parts of the wavefield, so we maintain their solution property. This implementation allowed for fast convergence of PINNs to the acoustic scattered wavefield solutions. To reduce the number of neurons needed in the Gabor layer, we used another, much smaller, auxiliary network to predict the Gabor apex/center location from the input coordinates. As a result, the Gabor center (apex) is a dynamic parameter that depends on the input samples, rather than a learned fixed parameter. The combination of these tools in this network resulted in an efficient PINN for learning wavefields. This efficiency was highlighted on realistic models for high-frequency wavefields.

\section*{Acknowledgements}
The authors thank KAUST for supporting this research. We would also like to thank the SWAG group for the collaborative environment. The data and accompanying codes that support the findings of this study will be available at: https://github.com/summerwine668/PINNbasedgabor.

% References
%\bibliographystyle{ieeetr}
\bibliography{references,paper}

\end{document}